\documentstyle[12pt]{article}
\input epsf

\begin{document}
\baselineskip=15pt
\newcommand{\xx}{{\bf x}}
\newcommand{\yy}{{\bf y}}
\newcommand{\del}{{\partial}}
\newcommand{\bc}{\begin{center}}
\newcommand{\ec}{\end{center}}
\newcommand{\be}{\begin{equation}}
\newcommand{\ee}{\end{equation}}
\newcommand{\bq}{\begin{eqnarray}}
\newcommand{\eq}{\end{eqnarray}}
\newcommand{\x}{{\bf x}}
\newcommand{\ILO}{I_{log}^{\circ \Lambda}}
\newcommand{\IQO}{I_{quad}^{\circ \Lambda}}
\newcommand{\p}{\varphi}
\newcommand{\CA}{{\cal{A}}}
\begin{titlepage}
\rightline{\large{\bf{hep-th/9811157}}}
\vskip1in
\begin{center}
{\Large{\bf Implicit Regularisation Technique: Calculation of the Two-loop
$\varphi^4_4$-theory $\beta$-function.}}
\end{center}
\vskip2.0cm
\begin{center}
{\large {\bf
A. Brizola} $^\dagger$, 
\bf{O. Battistel}$^\dagger$$^\ddagger$, {\bf Marcos Sampaio}$^\dagger$ and
{\bf M. C. Nemes}$^\dagger$ }\\
\vskip0.5cm
$\dagger$ Universidade Federal de Minas Gerais\\
Physics Department - ICEx\\
P.O.Box 702, 30161-970, Belo Horizonte - MG\\
Brazil \\
\vskip0.3cm
$\ddagger$ Universidade Federal de Santa Maria\\
Physics Department\\
P.O.Box 5093, 97119-900, Santa Maria -RS\\
Brazil\\

\vskip0.5cm

{\it msampaio@fisica.ufmg.br, brizola@fisica.ufmg.br, carolina@fisica.ufmg.br}
\end{center}
\vskip1in
%%%%%%%%%%%%%%%%%%%%%%ABSTRACT%%%%%%%%%%%%%%%%%%%%%%%%%%%%%%%
\begin{abstract}
\noindent
We propose an implicit regularisation scheme. The main advantage is that
since no explicit use of a regulator is made, one can in principle 
avoid  undesirable symmetry violations related to its choice. The divergent
amplitudes are split into basic divergent integrals which depend only
on the loop momenta and finite integrals. The former can be 
absorbed by a renormalisation procedure whereas the latter can be evaluated without
restrictions. We illustrate with the calculation of the $QED$ and $\varphi^4_4$-theory 
$\beta$-function to one and two-loop order, respectively.

\end{abstract}
%%%%%%%%%%%%%%%%%%%%%%%%%%%%%%%%%%%%%%%%%%%%%%%%%%%%%%%%%%%%%
\end{titlepage}

%%%%%%%%%%%%%%%%%%%%%%%%%%%%%%%%%%%%%%%%%%%%%%%%%%%%%%%%%%%%%

In dealing with ultraviolet divergences in perturbative calculations of Quantum
Field Theories (QFT), one is led to adopt a regularisation scheme (RS) to handle the
divergent
integrals. A vast arsenal of such schemes is presently available, viz. Dimensional
Regularisation (DR), Pauli-Villars (PV), Zeta-function Regularisation, Lattice
Regularisation, etc. . The choice of a particular scheme is
generally based on its adequacy to a particular computational
 task or compatibility with the underlying theory in the sense of preserving its
vital symmetries.
For example, DR is usually employed in particle physics since it
preserves unitarity and gauge invariance.  However, care must
be exercised in DR when parity-violating objects ($\gamma^5$-matrices,
$\epsilon_{\mu_1 \ldots \mu_n}$ tensors) occur in the theory \cite{CPM}. The 
properties of such objects depend very much on the space-time dimension 
and this clashes with the idea of analytic continuation on the dimension of the
space-time $D$. 
 
The issue of finding an ambiguity free RS so that the theory in consideration
is not plagued by RS-dependent amplitudes is most important, particularly in chiral
and non-renormalisable models. In the latter, the RS is frequently defined as a part
of the model. Consequently, any parameters introduced by a specific choice 
must be adjusted phenomenologically \cite{KLEVANSKY}, \cite{GHERGHETTA}. 

Recently a step in this direction has been taken. A technique was proposed
for the manipulation and calculation of divergent amplitudes in a way that a 
regularisation need only to be assumed {\bf implicitly} \cite{OAC},\cite{OCPRD}.
The main idea is to manipulate the integrands 
of the divergent amplitudes by means of  algebraic identities until the
physical content, i.e. the external momentum dependent part, is isolated
and displayed solely in terms of finite integrals \footnote{The philosophy is 
somewhat close in spirit to the BPHZ procedure \cite{BPHZ} where a Taylor expansion
is made around a fixed value of the physical momentum. In our approach any
identity which allows the amplitude to be written in the desired form may
be used (also Taylor expansion). In the case of amplitudes with different masses
the two philosophies become rather different as it was illustrated in \cite{OCPRD}.}.
 On the other hand, the
divergent content is automatically reduced to a set of basic divergent objects which
can be organised according to their degree of divergence. Throughout this process,
it is assumed that the ultraviolet divergent integrals in the momentum
(say, $k$) are regulated by the multiplication of the integrand by
 a regularising function $G( k^2, \Lambda_i)$,
\be
\int_{k} f(k) \rightarrow \int_{k} f(k) G(k^2, \Lambda_i^2) \equiv \int_{k}^\Lambda 
f(k) \, ,
\label{regulator}
\ee
$\int_{k} \equiv \int d^4 k/((2 \pi)^4)$ and
$\Lambda_i$ are the parameters of a distribution $G$ whose behaviour for large $k$
renders the integral finite. 

One important feature of  DR in what  concerns ambiguities is related
to the various possible choices for the momentum routing in amplitudes involving loops.
In this case there are correspondly as many amplitudes which, in principle,
 can be brought to the same form by adequate shifts in the integration variable. 
Whilst such shifts
are permitted in DR, for a $4$-D regularisation, if one effects a shift in the
integration variable, there should be a compensation by surface terms
as it is well known. This 
is precisely the origin of certain ambiguities and symmetry violations in many models
of physical interest \cite{GHERGHETTA}. Hence the question of how one should
proceed in situations beyond the scope of DR immediately arises. 
In this sense,  it was shown in \cite{OAC} that the same
consistency as exhibited  by DR regarding the momentum routing 
in the divergent integrals could be achieved in $4$-D regularisations
provided that a set of Consistency Relations (CR) which involve integrals
of the same degree of divergence were established, namely 
\bq
\int_{k}^\Lambda \frac{24 k_{\mu} k_{\nu} k_{\alpha} k_{\beta}}{(k^2-m^2)^4} &=&
(g_{\alpha \beta}g_{\mu \nu} + g_{\alpha \mu}g_{\nu \beta} + g_{\alpha \nu}g_{\mu
\beta}) I_{log}^\Lambda(m^2) \label{C1} \\
\int_{k}^\Lambda \frac{2 k_{\mu} k_{\nu}}{(k^2-m^2)^2} &=&
g_{\mu \nu}  I_{quad}^\Lambda(m^2) \label{C2} \\
\int_{k}^\Lambda \frac{4 k_{\mu} k_{\nu}}{(k^2-m^2)^3} &=&
g_{\mu \nu}  I_{log}^\Lambda(m^2) \label{C3}\, ,
\eq 
where
\bq
I_{log}^\Lambda(m^2) \equiv \int_{k}^{\Lambda} \frac{1}{(k^2 - m^2)^2} \label{Ilog}
\, ,\\
I_{quad}^\Lambda(m^2) \equiv \int_{k}^{\Lambda} \frac{1}{(k^2 - m^2)} \label{Iquad} \, .
\eq                                                         
The CR above are readily satisfied within the context of DR. 
It was also shown in \cite{OAC} that such CR can be obtained by demanding
the Green's functions of the theory to be translational invariant. In \cite{OCPRD}
the CR were proved to be the main ingredient in order to obtain unambiguous and
symmetry preserving amplitudes in the (gauged) Nambu-Jona-Lasinio model and thus
solving a long standing problem which has threatened its reliability. 
It is important to stress that the procedure adopted makes use solely of 
general properties of the regulator $G$ in (\ref{regulator})  avoiding an
explicit form. 

In this letter we consider the CR as the minimal
consistency conditions for $4$-D regularisations as a starting point. Then we show 
that although an explicit construction of regularising functions that fulfill the CR
can be made \cite{OAC}, one need not to do so. Instead, an ``implicit
regulator" is assumed and serves the purpose of mathematically  justifying 
the algebraic steps in the integrands of the divergent integrals. In the context 
of renormalisable theories such as $QED$ and $\varphi^4$-theory (to one and two loop
order, respectively), we show that a
renormalisation procedure in which the basic divergent integrals are absorbed
in the counterterms can be effected. This is an important check for
applications in the so called (super)renormalisable models \cite{CS}.
We illustrate with the calculation of
the renormalisation group $\beta$-function. 

The advantages of our formulation reside
in the fact that it provides a consistent RS which preserves important features of
DR yet being applicable where DR fails. Besides, the physical content of
the amplitudes will be displayed in terms of finite integrals only. This is of 
great value from the
phenomenological standpoint since regularisation prescriptions usually
modify the external momentum dependence and introduce non-physical behaviour such
as unitarity violation and unphysical thresholds.  From the aesthetical standpoint,
 it constitutes in a direct and economical RS that makes use solely of general
properties of the basic divergencies for which a regularisation needs only implicitly
to be assumed.                                     
\vskip0.5cm
%%%%%%%%%%%%%%%%%%%%%%%%%%%%%%%%%%%%% QED %%%%%%%%%%%%%%%%%%%%%%%%%%%%%%%%%%%%
We start with the calculation of the $QED$ $\beta$-function to one
loop order.
The renormalisation constants for the photon field, electron field, charge and mass
are defined as usual \cite{ZJ}, viz. $A^\mu_0 = Z_3^{1/2} A^\mu$,
$\Psi_0 = Z_2^{1/2} \Psi$, $e_0 = Z_1/(Z_2  Z_3^{1/2}) e$, $m_{e 0} =
(Z_0/Z_2) m_e$, with $Z_1/Z_2=1$, as required 
by gauge invariance. The Callan-Symanzik $\beta$-function can be written as
\cite{ZJ}:
\be
\beta_{QED} =  e \frac{\del [\ln Z_3^{1/2} (e_0,\Lambda/m)]}{\del \ln \Lambda} \, .
\label{betaqed}
\ee
The renormalisation constant $Z_3$ is calculated from the one loop correction
to the photon propagator namely the vacuum polarisation tensor
$\Pi_{\mu \nu}(q)$ whose amplitude reads
\bq
i \Pi_{\mu \nu} (q) &=& - e^2 \int^{\Lambda}_k 
\frac{Tr[\gamma_\nu (\gamma^\rho k_\rho - \gamma^\rho q_\rho + m_e) \gamma_\mu 
(\gamma^\rho k_\rho + m_e)]}{\Big[ (k-q)^2 - m_e^2\Big] (k^2 - m_e^2)} + \nonumber \\
&+& (Z_3 - 1) (q_\mu q_\nu - q^2 g_{\mu \nu}) \, ,
\label{VP}
\eq
where the last term on the RHS of (\ref{VP}) is the counterterm needed to absorb 
the divergence coming from the first to one loop order. Using the algebra of the Dirac
matrices and trace identities we find
\bq
i \Pi_{\mu \nu} (q) &=& -4 e^2 \{ 2 I_{\mu \nu}^\Lambda(m_e^2,q^2) -
 2 I_\mu^\Lambda (m_e^2,q^2)
 q_\nu - \frac{1}{2}
g_{\mu \nu} [ I_{quad}^\Lambda (m_e^2) + {\bar{I}}^\Lambda(m_e^2,q^2) \nonumber \\
&+& q^2 I^\Lambda (m_e^2,q^2) ] \} + (Z_3 - 1) (q_\mu q_\nu - q^2 g_{\mu \nu}) \, ,
\label{VP1}
\eq
where
\be
I_{\mu \nu}^\Lambda (m_e^2, q^2) = \int^{\Lambda}_k 
\frac{k_\mu k_\nu}{\Big[ (k-q)^2 - m_e^2\Big] (k^2 - m_e^2)} \, ,
\label{I1}
\ee
\be
I_{\mu}^\Lambda (m_e^2, q^2) = \int^{\Lambda}_k 
\frac{k_\mu}{\Big[ (k-q)^2 - m_e^2\Big] (k^2 - m_e^2)} \, ,
\label{I2}
\ee
\be
I^\Lambda (m_e^2, q^2) = \int^{\Lambda}_k 
\frac{1}{\Big[ (k-q)^2 - m_e^2\Big] (k^2 - m_e^2)} \, ,
\label{I3}
\ee
\be
{\bar{I}}^\Lambda(m_e^2, q^2) = \int^{\Lambda}_k 
\frac{1}{\Big[ (k-q)^2 - m_e^2\Big]} \, .
\label{I4}
\ee
Now we proceed to reorganise (\ref{VP1}) until it is reduced to (basic) divergent
integrals that depend only on the loop momenta. As a matter of
illustration let us take (\ref{I4}).
By using repeatedly one (possible) convenient algebraic identity 
at the level of the integrand, 
\be
\frac{1}{[(k-q)^2 - m_e^2]} = \frac{1}{(k^2 - m_e^2)} - \frac{q^2 - 2(k \cdot q)}
{(k^2 - m_e^2)[(k - q)^2 - m_e^2]} \, ,
\label{Identity}
\ee
until the divergent integrals carry no dependence on the external momentum $q$, enables
us to cast (\ref{I4}) as
\bq
\int^{\Lambda}_k  \frac{1}{\Big[ (k-q)^2 - m_e^2\Big]} &=& I_{quad}^\Lambda(m_e^2) - q^2
I_{log}^\Lambda(m_e^2) + q_\mu q_\nu \int_k^\Lambda \frac{4 k^\mu k^\nu}{(k^2-m_e^2)^3}
 + \nonumber\\ &+& \int_k \frac{q^4}{(k^2 -m_e^2)^3} - 
\int_k \frac{(q^2 - 2q \cdot k)^2}
{(k^2 - m_e^2)^2[(k-q)^2 - m_e^2]} \, .
\label{DCR}
\eq
The last two integrals above are finite and a calculation shows
that they cancel each other, whereas the CR (\ref{C3}) can be used to reduce
(\ref{DCR}) to $I_{quad}^\Lambda(m_e^2)$, as we expected. A similar procedure can
be employed in the other divergent integrals
(\ref{I1})-(\ref{I3}) using the CR (\ref{C1})-(\ref{C3}) and yields,
after a few algebra:
\be
i \Pi_{\mu \nu} = (q_\mu q_\nu - q^2 g_{\mu \nu}) \Bigg\{
 - \frac{4}{3 (4 \pi)^2} e^2 \Bigg[ \frac{q^2}{q^2 + 2 m_e^2}{\tilde{Z}}(m_e^2,q^2) 
+ \frac{1}{3} \Bigg] - \frac{4}{3}i e^2 I_{log}^\Lambda (m_e^2) + (Z_3 - 1) \Bigg\}
\, ,
\ee
where we defined
\be
{\tilde{Z}}(m^2,q^2) = \int_0^1 dz \ln \Big( \frac{q^2 z (1 - z) - m^2}{-m^2} \Big) \, ,
\label{Ztilde}
\ee
$z$ being a Feynman parameter. 
Finally we may choose the renormalisation constant such that
\be
(Z_3 -1) = \frac{4}{3}i e^2 I_{log}^\Lambda (m_e^2) \, ,
\label{Z3}
\ee
which, in this case, amounts to a subtraction at $q=0$. Before proceeding to
the calculation of the $\beta$-function, let us analyse the $\varphi^4_4$-theory
to two-loop order.
%%%%%%%%%%%%%%%%%%%%%%%%%%%%% END OF QED  %%%%%%%%%%%%%%%%%%%%%%%%%%%%%%%%%%%%

%%%%%%%%%%%%%%%%%%%%%%%%%%%%% PHI^4 %%%%%%%%%%%%%%%%%%%%%%%%%%%%%%%%%%%%%%%%%%
The $\varphi^4_4$-theory bare Lagrangian may written in terms of the 
renormalised parameters as
\be
{\cal{L}}_0 =
{\cal{L}}_R + {\cal{L}}_R^{CT} = \frac{1}{2} (\del_{\mu} \p)^2 - \frac{1}{2}
m^2 \p^2 - \frac{g}{4!} \p^4 
+ \frac{A}{2}  (\del_{\mu} \p)^2 - \frac{B}{2}
m^2 \p^2 - \frac{g C}{4!} \p^4 \, ,
\ee
where the set of variables $\p$, $m$, $g$ are related to the bare variables via the
renormalisation constants as $\p_0 = Z_\phi^{1/2} \p$, $m_0^2 = Z_m m^2$,
$g_0 = Z_g g$ and $Z_\phi = 1 + A$, $Z_m Z_\phi = 1 + B$, $Z_g Z_\p^2 = 1 + C$. Then
$A$, $B$ and $C$ are thought to have a series expansion in $g$, $A = \sum_{n=1}^\infty
a_n g^n$, etc., and define the
counterterms which cancel the infinities that emerge from the diagrammatic expansion
of the theory. Let us start with the one loop divergencies represented by the
``tadpole" and the ``fish" diagrams (figs.$1(a)$ and $1(d)$). 
%%%%%%%%%%%%%%%%%%%%%%%%%%%%%%%%%%%%%%%%%%%%%%%%%%%%%%%%%%%%%%%%%%%%%%%%%%%%%%%%%%%%%
\begin{figure}[h,t]
  \centerline{
    \epsfxsize=6in
    \epsffile{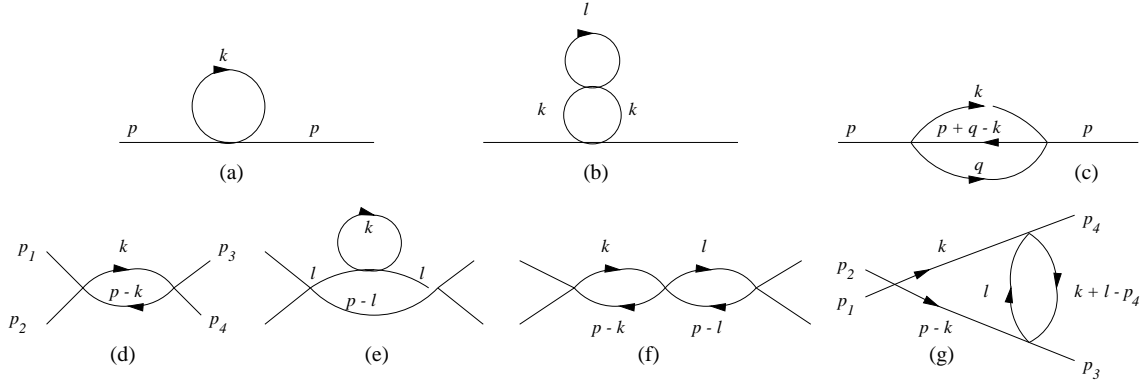}
             }
  \caption[diagrams]{\label{diagrams}$\varphi^4$-theory. Diagrams contributing
to the $2$ and $4$-point functions to $2$-loop order}  
\end{figure}
%%%%%%%%%%%%%%%%%%%%%%%%%%%%%%%%%%%%%%%%%%%%%%%%%%%%%%%%%%%%%%%%%%%%%%%%%%%%%%%%%%%%%

The Feynman rules applied to these diagrams, including the corresponding symmetry
factors, together with (\ref{Ilog}) and (\ref{Iquad}),  yield
\bq
- i \CA_a &=& \frac{(-i g)}{2} \int_k \frac{i}{k^2 - m^2} = \frac{g}{2}
I_{quad}^\Lambda (m^2) \, , \\
- i \CA_d &=& \frac{(-i g)^2}{2}  \int_k \frac{i^2}{(k^2 -
m^2)^2[(p-k)^2-m^2]} \label{Aa} \\ 
&=& \frac{g^2}{2} \Big( 3 I_{log}^\Lambda (m^2) - b \, [{\tilde{Z}}(m^2,s) + 
s \rightarrow t + s \rightarrow u] \Big) \label{Ad} \, ,
\eq
where $p^2 = (s,t,u)$ are the usual Mandelstam variables, $b = i/(4 \pi)^2$ and
${\tilde{Z}}(m^2,s)$ is defined as in (\ref{Ztilde}) .

In order to exhibit the 
divergences exclusively as a function of the
internal momentum (as we did  in (\ref{Ad}))
we make use of the identity (\ref{Identity}) in a similar fashion as it was done
for $QED$. Hence we may choose as counterterms 
\be
b_1 = -\frac{i}{2}\frac{I_{quad}^\Lambda (m^2)}{m^2} \quad\quad , \quad\quad 
c_1 = -\frac{3 i}{2} I_{log}^\Lambda(m^2) 
\ee
whilst $a_1 = 0$. Now we proceed to $2$-loop order. As it is well 
known, the $1$-loop counterterms must be taken to higher orders to cancel the 
divergencies of the corresponding subdiagrams. The ``double-scoop" diagram (fig.$1b$) 
represents the amplitude
\be
- i \CA_b =  \frac{(-i g)^2}{4} \int_{k,l} \frac{i^3}{(k^2 - m^2)^2(l^2-m^2)} 
= \frac{i g^2}{4} I_{quad}^\Lambda (m^2) I_{log}^\Lambda (m^2) 
\ee
%%%%%%%%%%%%%%%%%%%%%%%%%%%%%%%%%%%%%%%%%%%%%%%%%%%%%%%%%%%%%%%%%%%%%%%%%%%%%%%%%%%%%
\begin{figure}[h,t]
  \centerline{
    \epsfxsize=4in
    \epsffile{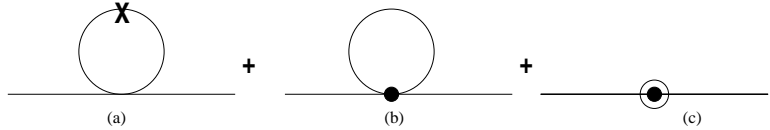}
             }
  \caption[Counteterms]{\label{ctterms}Counterterm diagrams for the ``double-scoop"}  
\end{figure}
%%%%%%%%%%%%%%%%%%%%%%%%%%%%%%%%%%%%%%%%%%%%%%%%%%%%%%%%%%%%%%%%%%%%%%%%%%%%%%%%%%%%%
Now we must take into account the counterterm diagrams represented in fig.$2$. The
first two correspond to the mass and vertex counterterms, respectively. The third is
the counterterm associated with diagram $2(b)$ which exactly cancels it. The sum of 
all the contributions vanish, as they should, since the ``double scoop" diagram does
not depend on the external momentum and hence it generates a purely divergent
contribution.
\vskip0.4cm
Since the integration over the internal loop momenta $k$ and $l$
factorise, the $2$-loop scaterring amplitude depicted in fig.$1(f)$ can be written 
as
\be
-i \CA_f = \frac{i g^3}{4} \Bigg( \int_k \frac{1}{(k^2-m^2)[(k-q)^2-m^2]} \Bigg)^2 =
 \frac{i g^3}{4} \Big(I_{log}^\Lambda(m^2) - b {\tilde{Z}}(m^2,q^2) \Big)^2 \, .
\label{Af}
\ee
It is easy to see that the crossed term in (\ref{Af}) is exactly 
cancelled by the $1$-loop
counterterm associated with the graph $1(f)$. Consequently this graph
gives rise to a new counterterm $c_2$ proportional to
 $(I_{log}^\Lambda(m^2))^2$. 

Now, since the $1$-loop mass
counterterm cancels out graph $1(e)$, the only other contribution to $c_2$ comes from
graph $1(g)$:
\be
-i \CA_g = \frac{(-i g)^3}{4} \int_{k,l} \frac{i^4}{(k^2-m^2)(l^2-m^2)[(p-k)^2-m^2]
[(k+l-p_4)^2-m^2]} \, ,
\label{Ag}
\ee
$p^2 = (s,t,u)$. 
Notice, however, that the loop integrations do not factorise (overlapping divergence).
Within our strategy, we expect to be able to define other objects than (\ref{Ilog}) and
(\ref{Iquad}). Thus we manipulate (\ref{Ag}) using (\ref{Identity}) repeatedly to
finally obtain
\bq
-i \CA_g &=& \frac{i g^3}{4} \Big( 3 \ILO (m^2) + F_{fin}(p_i,p^2) \Big) \, , \\
\ILO (m^2) &\equiv& \int_{k,l}^\Lambda \frac{1}{(k^2-m^2)^2[(k+l)^2-m^2](l^2-m^2)} 
\label{Ilogo} \, ,
\eq
where $F_{fin}(p_i,p^2)$ is a finite function of the external momenta $p_i$, $p^2$ and
(\ref{Ilogo}) is our new basic (logarithmically) divergent object. So
we can write
\be
c_2 = \frac{3}{4} \Big( (I_{log}^\Lambda(m^2))^2 + \ILO (m^2) \Big) \, .
\ee

We are now left with the ``setting sun" diagram depicted in fig.$1(c)$ whose
contribution to the $2$-point function is given by
\be
-i \CA_c = \frac{(-i g)^2}{6} \int_{k,q}\frac{i^3}{(k^2-m^2)[(p+q-k)^2-m^2](q^2-m^2)}
\, .
\label{Ac}
\ee
Again the loop integrations do not factorise. Notwithstanding, a recursive use
of relation (\ref{Identity}) together with the CR (\ref{C3}) enable us to expand
(\ref{Ac}) and write, after
some algebra, 
\bq
\CA_c &=& -\frac{g^2}{6} \IQO (m^2) + \frac{g^2 b}{12} p^2 I_{log}^\Lambda (m^2) + g^2
G_{finite} (p^2) \\
\IQO &=& \int_{k,l}^\Lambda \frac{1}{(k^2-m^2)(l^2-m^2)[(l-k)^2-m^2]} \, ,
\label{Iquado}
\eq
where (\ref{Iquado}) is another basic divergent object
with overlapping quadratic divergence. Hence we have the two remaining
counterterms, namely
\be
a_2 = \frac{b}{12} I_{log}^\Lambda (m^2) \,\,\, , \,\,\, b_2 = \frac{1}{6}
\frac{\IQO (m^2)}{m^2} \, .
\ee

\vskip0.5cm

Having obtained our basic divergent objects to $2$-loop order, some of their
properties will be useful. They can be related to each other in a simple fashion.
By differentiating these objects with respect to (squared) mass  one makes them more 
convergent. It can be seen from (\ref{Ilog}),  (\ref{Iquad}),  (\ref{Ilogo}) and
(\ref{Iquado}) that
\be
\frac{\del I_{log}^\Lambda (m^2)}{\del m^2} = \frac{-b}{m^2} \,\, ,
\frac{\del I_{quad}^\Lambda (m^2)}{\del m^2} = I_{log}^\Lambda (m^2) \,\, ,
\frac{\del \ILO (m^2)}{\del m^2} = \frac{\eta}{m^2} \,\, ,
\frac{\del \IQO (m^2)}{\del m^2} = 3 \ILO (m^2) \,\, ,
\label{DerI}
\ee
($\eta = -1/(96 \pi^4)$).

To test our results we can calculate the  $\beta$-function to
$O(\hbar^2)$. In this case the $\beta$-function can be written as
\be
\beta_{\varphi^4}= - g \frac{\del [\ln {\bar{Z}}(g_0,\Lambda/m)]}{\del \ln \Lambda} \, ,
\label{betaphi4}
\ee
where ${\bar{Z}} = Z_g^{-1} Z_{\p}^2$.

It follows from dimensional analysis
that the argument of our logarithmically divergent objects is $m^2/\Lambda^2$.
Using the counterterms $a_1$, $a_2$, $c_1$ and $c_2$ which we have calculated
and the relations (\ref{DerI}), we obtain the two first well-known coefficients
of the $\varphi^4_4$-theory $\beta$-function:
\be
\beta_{\varphi^4} = \frac{3}{16 \pi^2} g^2 - \frac{17}{768 \pi^4} g^3 \, .
\ee
%%%%%%%
Similarly the $\beta$-function of $QED$ can be calculated using (\ref{betaqed}),
  (\ref{Z3}) and  (\ref{DerI}) to give the well-known result:
\be
\beta_{QED} = e^3/(12 \pi^2) \, .
\ee
%%%%%%%

\vskip0.5cm

To conclude: We have tested an implicit regularisation scheme by calculating
the renormalisation group $\beta$-function. Its main ingredient is a set
of Consistency Relations which relate integrals of the same degree of divergence
(\ref{C1}), (\ref{C2}), (\ref{C3}).
 Since we do not resort to a specific
regulator, we believe that it can be a useful tool to revisit relevant ambiguity
problems regarding RS choice in both renormalisable and non-renormalisable $4$-D QFT.
Although we have presented a $4$-D formulation, this framework can be extended 
to arbitrary dimensions. In particular, in $3$-D it can be useful to deal with
RS ambiguities in the Chern-Simons-Matter models
 \cite{CS}, \cite{WIP}.

\vskip0.4cm

The authors wish to thank Dr. O. Piguet for useful discussions.
MS and MCN acknowledge a grant from CNPq/Brazil and AB from FAPEMIG/Brazil.

\end{document}